\documentclass[english,12pt]{article}
\usepackage[LGR,T1]{fontenc}
\usepackage[latin9]{inputenc}
\usepackage[letterpaper]{geometry}
\usepackage{babel}
\usepackage{amsmath}
\usepackage{graphicx}
\usepackage{setspace}
\usepackage{esint}
\usepackage[unicode=true,
 bookmarks=false,
 breaklinks=false,pdfborder={0 0 0},backref=false,colorlinks=false]
 {hyperref}
\hypersetup{pdftitle={Trade arrival dynamics and quote imbalance in a limit order book},
 pdfauthor={Lipton, Pesavento, Sotiropoulos}}

\makeatletter

\DeclareRobustCommand{\greektext}{%
  \fontencoding{LGR}\selectfont\def\encodingdefault{LGR}}
\DeclareRobustCommand{\textgreek}[1]{\leavevmode{\greektext #1}}
\DeclareFontEncoding{LGR}{}{}
\DeclareTextSymbol{\~}{LGR}{126}
\providecommand{\tabularnewline}{\\}

\newcommand{\phimax}{\ensuremath{\varpi}}

\usepackage{verbatim}

\makeatother

\begin{document}

\title{Trade arrival dynamics and quote imbalance in a limit order book}

\begin{singlespace}

\author{{\footnotesize{Alexander Lipton}}%
\thanks{{\footnotesize{Bank of America Merrill Lynch and Imperial College}}%
}{\footnotesize{, Umberto Pesavento}}%
\thanks{{\footnotesize{Bank of America Merrill Lynch}}%
}{\footnotesize{ and Michael G Sotiropoulos}}%
\thanks{{\footnotesize{Bank of America Merrill Lynch}}%
}}
\end{singlespace}

\date{2 December 2013}
\maketitle
\begin{abstract}
We examine the dynamics of the bid and ask queues of a limit order
book and their relationship with the intensity of trade arrivals.
In particular, we study the probability of price movements and trade
arrivals as a function of the quote imbalance at the top of the limit
order book. We propose a stochastic model in an attempt to capture
the joint dynamics of the top of the book queues and the trading process,
and describe a semi-analytic approach to calculate the relative probability
of market events. We calibrate the model using historical market data
and discuss the quality of fit and practical applications of the results. 
\end{abstract}

\section{Introduction}

The prevalence of computer driven trading has radically changed the
market structure in several asset classes, most notably in equities
and futures. This has attracted the interest of researchers driven
by both practical and academic motivations. Early studies have followed
an econometric approach in exploring the relationship between order
flow and prices within electronic trading venues \cite{Hasbouck1991,Engle2000,bouchaud2002,Smith2003},
whereas later studies focused on understanding and reducing the price
impact caused by large orders \cite{Almgren2005,Bouchaud2008}. More
recently, several researchers have focused on the application of point
processes \cite{Cont2010b,bacry2013}, queueing theory \cite{Cont2011}
and agent's utility functions \cite{Avellaneda2010} to modeling electronic
markets.

Participants in electronic trading fall into several categories, with
different characteristics and objectives. Although any formal classification
is blurred by the fact that the same market participants often interact
in several capacities at the same time, algorithmic traders can be
roughly divided into the following categories: market makers, who
provide the most liquidity in the markets and try to capture the bid-ask
spread with minimum duration risk; systematic traders and arbitrageurs,
who typically try to profit from price dislocations and statistical
relationships among the prices of different securities; and agency
brokers, who execute large trades on behalf of their customers. Irrespective
of their objectives, all market participants in a public electronic
venue contribute to price formation by adding and removing liquidity
in a limit order book where quotes are published and orders are matched.

Here, we will analyze the dynamics of prices and trades from an agency
broker's point of view, faced with the task of buying or selling a
given number of shares within a set time horizon and at the best possible
price. The broker typically seeks to obtain liquidity from a variety
of market venues, including public exchanges, dark pools and internal
liquidity sources. In a public exchange, a broker has the choice to
either remove liquidity from the far side (that is, the ask side for
a buy order or the bid side for a sell order) or add liquidity to
the near side of the order book. Either strategy has its advantages
and trade-offs. When removing liquidity, one pays the full spread
but is free to choose the timing of the trade. When posting passively,
one avoids paying the spread but gives up the timing option. In order
to explore this trade-off, we study the relationship between price
dynamics and intensity of trade arrival. In the next sections we will
first compute empirically average price moves and waiting times conditional
on the state of the order book. We then attempt to capture the main
features by introducing a stochastic model for diffusion in three
dimensions. We compute the probabilities of price movement and trade
occurrence from the model, and calibrate them to recent historical
market data.

\begin{center}
\begin{figure}
\begin{centering}
\includegraphics[scale=0.4]{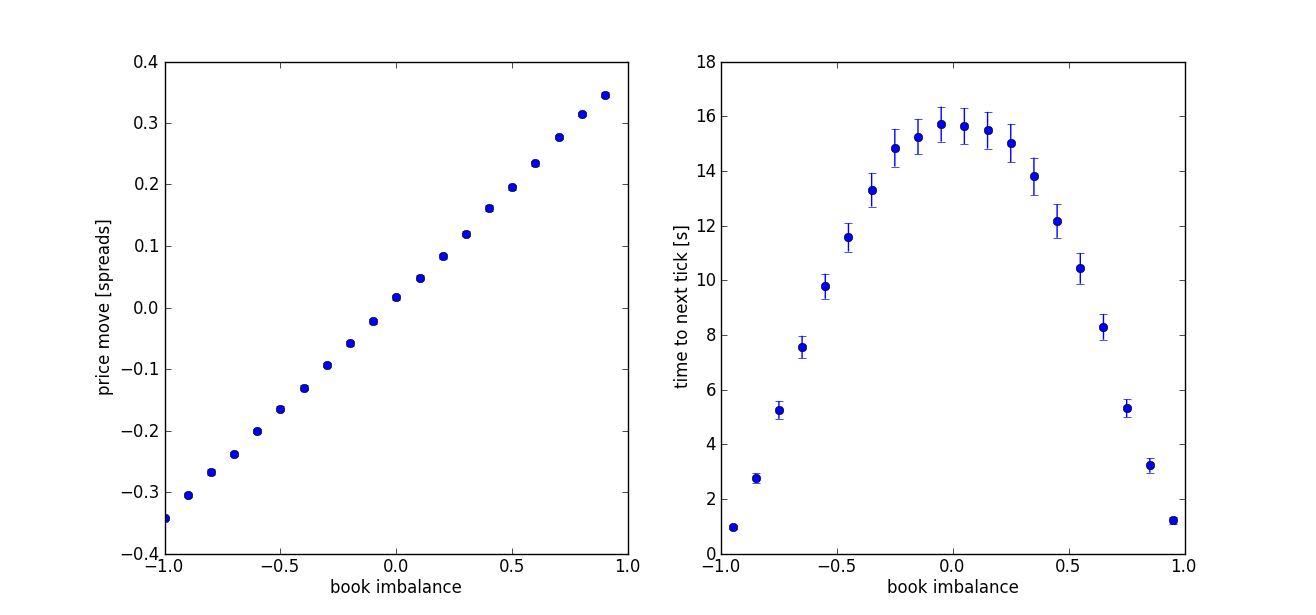}
\par\end{centering}

\caption{Average mid price move normalised by the bid-ask spread and waiting
time until the next mid price move as a function of book imbalance.
The data display clearly the non-martingale nature of prices at the
short time scales considered here. In the case shown, the average
price move can be up to a third of the spread in a highly imbalanced
book. The data shown are obtained by averaging the mid price moves
of VOD.L for all trading days in the first quarter of 2012. The data
are bucketed according the initial imbalance in the book and averaged
over the entire trading day. Error bars are calculated by treating
the daily averages of price variations as independent data points.
The daily average variations display error bars so small that they
are hardly visible in the scale of the plot above. \label{fig:cond_imbalance_emp-1}}
\end{figure}

\par\end{center}

\begin{table}
\begin{tabular}{ccccc}
$t_{0}$ & $p_{0}^{b}$ & $p_{0}^{a}$ & $q_{0}^{b}$ & $q_{0}^{a}$\tabularnewline
$t_{1}$ & $p_{1}^{b}$ & $p_{1}^{a}$ & $q_{1}^{b}$ & $q_{1}^{a}$\tabularnewline
$t_{2}$ & $p_{2}^{b}$ & $p_{2}^{a}$ & $q_{2}^{b}$ & $q_{2}^{a}$\tabularnewline
$\tilde{t_{0}}$ & $\tilde{p_{0}}$ & $\tilde{q_{0}}$ & $\tilde{s_{0}}$ & \tabularnewline
$t_{3}$ & $p_{3}^{b}$ & $p_{3}^{a}$ & $q_{3}^{b}$ & $q_{3}^{a}$\tabularnewline
$t_{4}$ & $p_{4}^{b}$ & $p_{4}^{a}$ & $q_{4}^{b}$ & $q_{4}^{a}$\tabularnewline
$\tilde{t_{1}}$ & $\tilde{p_{1}}$ & $\tilde{q_{1}}$ & $\tilde{s_{1}}$ & \tabularnewline
$t_{5}$ & $p_{5}^{b}$ & $p_{5}^{a}$ & $q_{5}^{b}$ & $q_{5}^{a}$\tabularnewline
$t_{6}$ & $p_{6}^{b}$ & $p_{6}^{a}$ & $q_{6}^{b}$ & $q_{6}^{a}$\tabularnewline
$\tilde{t_{2}}$ & $\tilde{p_{2}}$ & $\tilde{q_{2}}$ & $\tilde{s_{2}}$ & \tabularnewline
... & ... & ... & ... & ...\tabularnewline
\end{tabular}

\caption{Price moves and waiting times until trade arrival. We denote quote
updates by tuples $(t_{i},p_{i}^{b},p_{i}^{a},q_{i}^{b},q_{i}^{a})$,
with the quantities indicating respectively the time of the quote
update, the new best bid, the new best ask, the number of shares at
the best bid and those at the ask. The change in just one of the prices
or quantities is sufficient to trigger the publication of a quote
update, i.e. a new row in the table. In the same table we denote a
trade execution by the tuple $(\tilde{t_{j}},\tilde{p_{j}},\tilde{q_{j}},\tilde{s_{j}})$
with quantities indicating the time, the price, the quantity and the
side of the trade. This sequence is used for computing price moves
and waiting times until the arrival of the next trade, as described
in the text\label{tab:taq-proc}}
\end{table}

\section{Empirical observations}

A common intuition among market practitioners is that the order sizes
displayed at the top of the book reflect the general intention of
the market. When the number of shares available at the bid exceeds
those at the ask, participants expect the next price movement to be
upwards, and inversely, for the ask. This is why, for example, a broker
working an order on behalf of a client might be concerned about posting
too much of it at the near side, thus showing his intentions to the
rest of the market. In order to quantify and model this intuition,
we calculate a few quantities related to the microstructure of the
order book conditional on its bid-ask imbalance defined as 
\begin{equation}
I=\frac{q^{b}-q^{a}}{q^{b}+q^{a}},\label{eq:imbaldef}
\end{equation}
where $q^{b}$ and $q^{a}$ are the bid and ask quantities posted
at the top of the book. Positive (negative) imbalance indicates an
order book that is heavier on the bid (ask) side. In Figure \ref{fig:cond_imbalance_emp-1},
we show the effect of the book imbalance on the average mid price
change and on the waiting time until the next price change. We calculate
these quantities by considering the stopping time defined by the next
change in either the best bid or the best ask. The average price changes
are therefore related to the conditional up-tick and down-tick probabilities,
which have been previously considered in the literature \cite{Cont2011}.
Here, we use the same probabilities to compute the average size of
the price jump%
\footnote{A mid price change event can be induced by several actions, such as
a trade, a cancellation or even the addition of a new quote between
the current bid and ask spread, if the spread is big enough. None
of these actions is a sufficient condition for the mid price to change,
but at least one of them is necessary.%
}. As expected, a high book imbalance is indicative of the general
trading intention in the market and, on average, a good predictor
of mid price movements. The price change until the next tick is well
approximated by a linear function of the imbalance and is typically
well below the bid-ask spread, even for highly imbalanced order books.
In other words, although book imbalance might be used as a predictor
for the next price movement, it does not by itself offer an opportunity
for a straightforward statistical arbitrage.

The average waiting time until the next mid price change is also affected
by the imbalance of the order book, with highly imbalanced books indicating
a price move coming in a relatively short time. This relationship
is to be expected in a market where the typical broker posts part
of his orders at the near side and both price variations and queue
levels are mostly driven by the pressure of the order flow. In the
context of microstructure studies, this is usually interpreted as
the order flow having an impact on the limit book, but it can also
be seen more generally as the natural supply and demand influence
on the price of an asset. 

Another stopping time with economic significance is the one defined
by the first arrival of a trade of a specified side. This stopping
time is important to a broker who is posting part of his order at
the near side of the limit order book and has to wait for market orders
originating from the opposite side to be matched with his resting
order. Trades originating from the same side (i.e. trades from buy
market orders while the broker is posting at the bid) may affect the
share price and change the prevailing dynamics in the market, but
they will not contribute to the broker's fill rate. Therefore, for
an order posted at the bid (ask) side, the relevant stopping time
is the time of first arrival of a sell (buy) trade.

In Table \ref{tab:taq-proc} we show a typical segment of the trades
and quotes times series. It consists of quote tuples $(t_{i},p_{i}^{b},p_{i}^{a},q_{i}^{b},q_{i}^{a})$,
indicating the time of the quote update, the new best bid and ask
prices and sizes, and trade tuples $(\tilde{t_{j}},\tilde{p_{j}},\tilde{q_{j}},\tilde{s_{j}})$
indicating the time, price, quantity and side of the trade. Note that
although the side of the trade is not usually published by equities
exchanges, it can be inferred by its price compared to the prevailing
quotes. We calculate average price variations and waiting times until
the next buy or sell trade as follows: a) for each quote $(t_{i},p_{i}^{b},p_{i}^{a},q_{i}^{b},q_{i}^{a})$
we compute the prevailing book imbalance $I$, as in eq. (\ref{eq:imbaldef});
b) given a quote, we identify the next buy and sell trade on the tape;
c) we determine the prevailing mid price at the time of the trade;
d) we compute the difference between the prevailing mid price and
the mid price of the original quote, and assign it to the corresponding
bucket based on the original quote imbalance and the side of the trade.
A similar procedure is used for computing the waiting time until the
next buy or sell trade.

\begin{figure}
\begin{centering}
\includegraphics[scale=0.4]{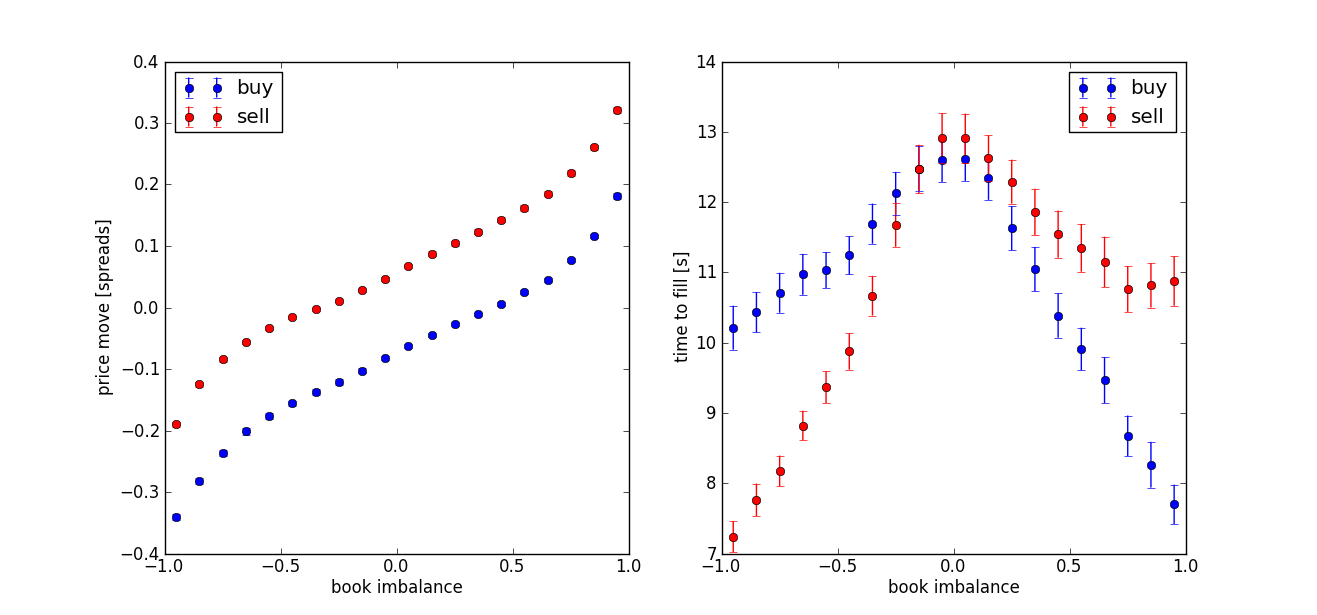}
\par\end{centering}

\caption{Mid-price moves normalised by the bid-ask spread and waiting times
until arrival of the next buy or sell trade. The average mid-price
moves and waiting times are shown as a function of the book imbalance.
Again, the expected price move displays a strong dependence on the
book imbalance. However it also shows a clear dependence on the side
of the trade determining the stopping time. \label{fig:cond_imbalance_emp-2}}
\end{figure}

\begin{figure}
\begin{centering}
\includegraphics[scale=0.4]{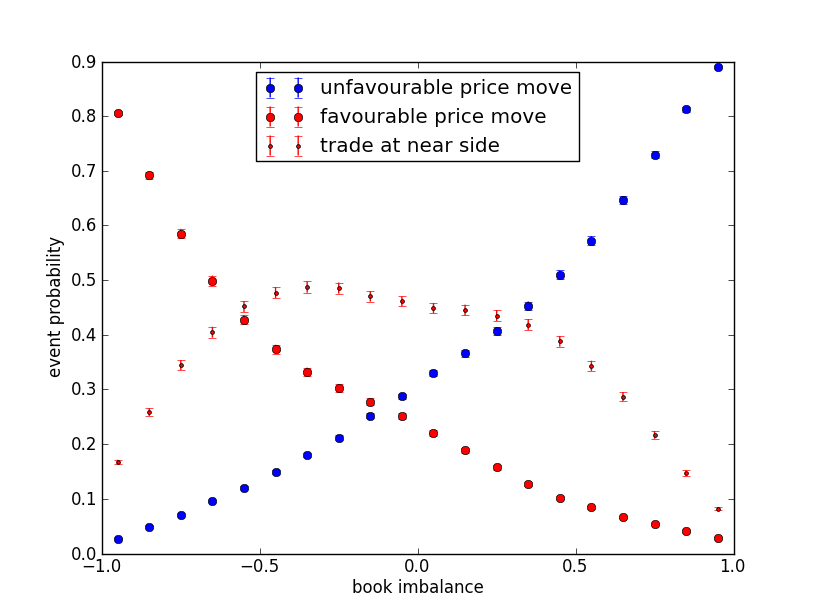}
\par\end{centering}

\caption{Probabilities of market events as a function of the book imbalance.
The data are obtained by averaging probabilities of market events
bucketed by the imbalance in the order book for VOD.L for all trading
days in the first quarter of 2012. Error bars are obtained by treating
averages of different trading days as independent data points. Favourable
or unfavourable price moves are from the point of view of a broker
who is posting part of his order at the near side of the order book.
\label{fig:cond_prob_empir-1}}
\end{figure}

In Figure \ref{fig:cond_imbalance_emp-2} we show average price moves
and waiting times until the arrival of the next buy or sell trade,
conditional on the book imbalance. As expected, the general trends
of these quantities as a function of the book imbalance are similar
to those obtained by using the next mid price move as a stopping time.
However average price movement from the observation time to the arrival
of the next sell trade display a clear shift upwards compared to those
derived until the arrival of the next buy trade. This can be interpreted
as the result of the information advantage of aggressive traders over
traders posting at the near side. It can also be interpreted more
mechanically, as the impact of the intervening buy trades between
the observation time and the next sell trade (conversely, the impact
of intervening sell trades if the stopping time is the next buy trade).
As a result of this information, a trader who is posting his quote
on the bid side of the limit order book will see on average an upwards
price move by the time a sell trade matches his quote. Conversely,
a trader posting on the ask side of a book displaying the same book
imbalance will experience a price movement with a downward bias. Finally,
in Figure \ref{fig:cond_prob_empir-1} we compute the empirical probability
of the next market event as a function of the current book imbalance.
The events of interest are a favourable or unfavourable price move
or the occurrence of a matching trade from the point of view of a
broker pegging part of his order on the near side, i.e. the broker
keeps updating his quote to stay at the prevailing best bid or ask
price.

In the next sections, we will explain the empirical observations summarized
in Figures \ref{fig:cond_imbalance_emp-1} and \ref{fig:cond_prob_empir-1}
by introducing a three-dimensional stochastic model for the joint
evolution of the bid/ask queues and the near side trade arrival process.
Before developing the full model we will first review a simpler, two-dimensional
model for the bid and ask queues only, in order to introduce some
of the analytical techniques that are used in the full three-dimensional
model.

\section{Modeling the bid and ask queues\label{sec:Modeling}}

Our starting point is the two-dimensional diffusion model for the
top of the book \cite{Cont2011}.We model the number of shares $q^{b}$
and $q^{a}$ posted at the top of the order book for the bid and ask
with the following stochastic processes:

\begin{equation}
(dq^{b},dq^{a})=(dw^{b},dw^{a}),\label{eq:b}
\end{equation}
where $w^{b},\, w^{a}$ are Brownian motions. Because the number of
shares available at the top of the book is always a positive quantity,
when either of the two processes cross the positive axis we reset
its value by drawing from two positive distributions $q_{initial}^{b}$
and $q_{initial}^{a}$. One can think of these distributions as modeling
the next price levels in the book; once the first level is depleted,
they serve as a new starting point for the top queue. Following this
interpretation, we will also assume that every time one of the two
queues is depleted the price also moves in the direction of the depleted
queue. If the ask queue is depleted the price moves up; if the bid
is depleted the price moves down. In other words, we assume that as
one side of the book is depleted under the pressure of incoming aggressive
trades and cancellations of existing limit orders, the other side
follows it to keep the bid-ask spread fixed. 

Given the queue model of equation (\ref{eq:b}), it is natural to
ask whether an up-tick or a down-tick is the most likely future price
move. We can calculate such probabilities by identifying the $x$
and the $y$ axes of the plane with the bid and the ask quote sizes
respectively and writing the general equation for the evolution of
the hitting probability:

\begin{equation}
\frac{1}{2}P_{xx}+\frac{1}{2}P_{yy}+\rho_{xy}\ P_{xy}=0,\label{eq:pde2dxy}
\end{equation}
where $\rho_{xy}$ is the correlation between the diffusion processes
governing the depletion and replenishment of the bid and ask queues.
This correlation is typically negative in a normal market. The choice
of boundary conditions selects the event with corresponding probability
$P$. For example, for an up-tick price movement, the boundary conditions
are
\begin{equation}
P\left(x,0\right)=1,\quad P\left(0,y\right)=0.\label{eq:xy_upmove_bc}
\end{equation}
Moreover, since the probability $P$ is for event occurrence up to
the first stopping time, the PDE in eq. (\ref{eq:pde2dxy}) is time
independent. In fact, the solution is the $T\rightarrow\infty$ limit
of the fixed $T$ horizon problem.

As shown in \cite{Lipton2001,Cont2011,Lipton2013} among others, it
is possible to solve the above PDE by introducing two changes of variables.
The first transformation removes the correlation between the queue
processes:

\begin{equation} \left\{ \begin{aligned} \displaystyle \alpha(x,y) =\ & \displaystyle x\\ \displaystyle \beta(x,y) = & \displaystyle \frac{\left(-\rho_{xy} x + y\right)}{\sqrt{1-\rho^2_{xy}}} , \end{aligned} \right. \label{eq:2D_1stVars} \end{equation}yielding
the equation

\begin{equation}
P_{\alpha\alpha}+P_{\beta\beta}=0.
\end{equation}
The second transformation casts the modified problem in polar coordinates:

\begin{equation} \left\{ \begin{aligned} \displaystyle \alpha =& \displaystyle r \sin(\varphi)\\ \displaystyle	\beta =& \displaystyle r \cos(\varphi) \end{aligned} \right. \longleftrightarrow \left\{ \begin{aligned} r =& \sqrt{\alpha^2+\beta^2}\\ \varphi =&  \textrm{arctan}\left(\frac{\alpha}{\beta}\right), \end{aligned} \right. 	\label{eq:2D_2ndVars} \end{equation}where
$\cos\phimax=-\rho_{xy}$. Then the equation for the hitting probability
becomes

\begin{equation}
P_{\varphi\varphi}(\varphi)=0,
\end{equation}
and the boundary conditions for an up-tick price movement become
\begin{equation}
P\left(0\right)=0,\quad P\left(\phimax\right)=1.\label{eq:ab_upmove_bc}
\end{equation}
 The solution in polar coordinates is straightforward, $P(\varphi)=\varphi/\phimax$,
which in the original set of coordinates has the form

\begin{equation}
P(x,y)=\frac{1}{2}\left(1-\frac{\arctan(\sqrt{\frac{1+\rho_{xy}}{1-\rho_{xy}}}\frac{y-x}{y+x})}{\arctan(\sqrt{\frac{1+\rho_{xy}}{1-\rho_{xy}}})}\right).
\end{equation}
This is the probability of upward movement of a Markovian order book
in the diffusive limit.

\section{Adding trade arrival dynamics\label{sec:Adding-trade-arrival}}

In order to capture the joint dynamics of the bid and ask queues and
trade arrival, we introduce another stochastic process to model the
arrival of trades on the near side of the book (the bid side for a
broker executing a buy order, or the ask side for a sell order):

\begin{equation}
(dq^{b},dq^{a},d\phi)=(dw^{b},dw^{a},dw^{\phi})\label{eq:T}
\end{equation}
The process $\phi$ does not correspond to a market observable. However,
in analogy to our model for the bid and ask queues, we assume that
a new trade hits the near side whenever $\phi$ crosses the origin.
Within this extended three dimensional model, the queue processes
$q^{b}$ and $q^{a}$ are now driven exclusively by the addition and
cancellation of limit orders in the book until the arrival of a trade,
while the timing of near side trade arrivals is governed by the $\phi$
process. More importantly, we can model the relationship between book
imbalance, order flow and price variations by introducing correlations
between the stochastic processes $dw^{b}$, $dw^{a}$, and $dw^{\phi}$. 

The introduction of an unobserved diffusion process $dw^{\phi}$ for
modeling trade arrival leads to a non-Markovian model. As in the two-dimensional
case, we assume that once a queue gets depleted or a trade arrives,
the process $w^{\phi}$ restarts at a value $\phi_{0}$ which characterizes
the trade arrival time distribution. This model parameter is to be
determined by calibration.

During the dynamic optimization of an execution schedule, one needs
to consider the relative likelihood of favourable and adverse price
moves and of trade arrival at the near side of the book. In terms
of the model described above, these events correspond to the three
dimensional stochastic process $(q^{b},q^{a},\phi)$ crossing the
positive orthant. In the following, we calculate the probability of
these market events conditional on the state of the queue and the
value of the trade process $\phi$. As in the previous section we
identify the $x$ and $y$ coordinates with the processes governing
the bid and ask queue sizes, and the $z$ coordinate with the level
of the process $\phi$. The equation for the hitting probability becomes:
\begin{multline}
\frac{1}{2}P_{xx}+\frac{1}{2}P_{yy}+\frac{1}{2}P_{zz}+\rho_{xy}P_{xy}+\rho_{xz}P_{xz}+\rho_{yz}P_{yz}=0,
\end{multline}
 where $\rho_{xy}$ is the correlation between the processes governing
the bid and ask queues, and $\rho_{xz}$ and $\rho_{yz}$ are the
correlation of those processes with that governing the arrival of
the trades at the near side. The different events are identified by
the boundary conditions of $P$, which are set to one on the plane
corresponding to the market event considered, and zero on all other
boundaries. For example, the probability of a near side trade before
any price move corresponds to the boundary conditions
\begin{equation}
P\left(x,0,z\right)=0,\quad P\left(0,y,z\right)=0,\quad P\left(x,y,0\right)=1.\label{eq:xyz_trade_bc}
\end{equation}

\begin{figure}
\centering{}\includegraphics[scale=0.75]{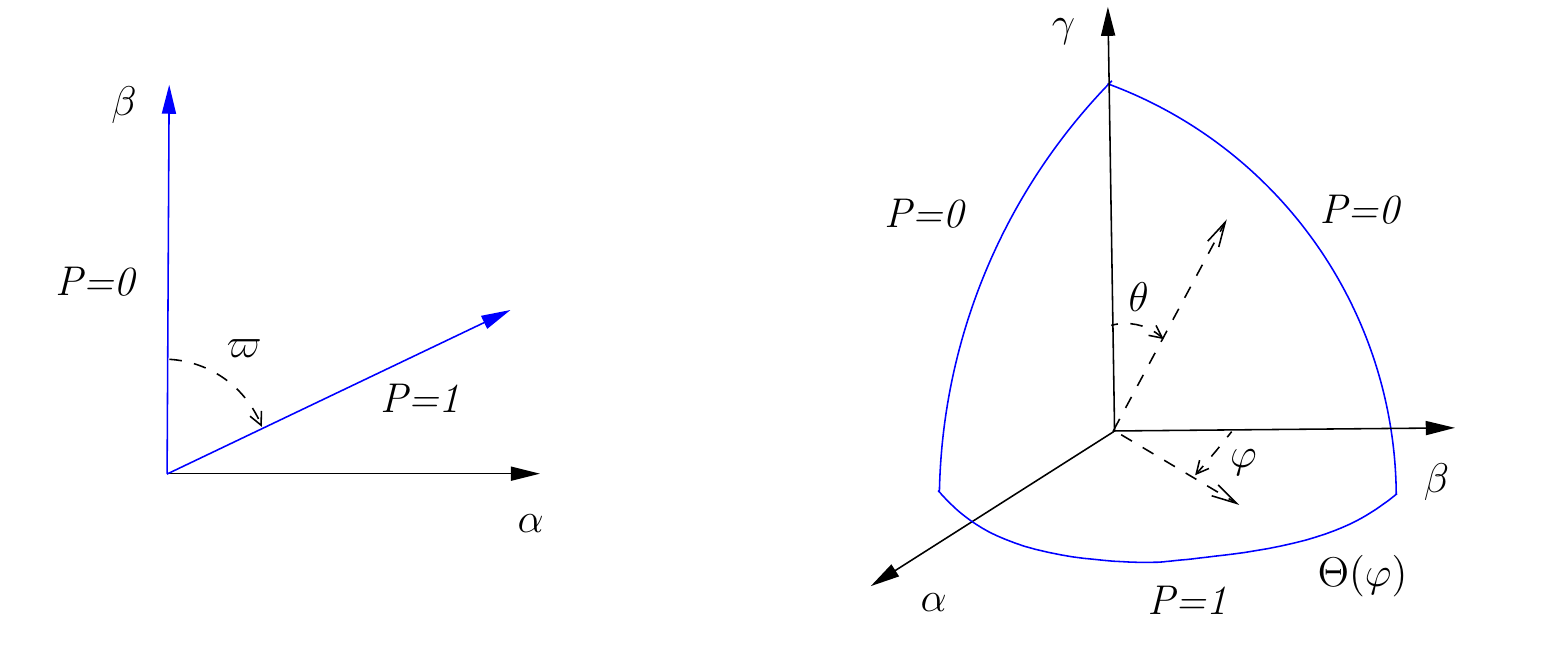}\caption{Integration domains for the two-dimensional (left) and three-dimensional
model (right). The boundary conditions shown correspond to an up-tick
price movement in the two-dimensional case, and to a near side trade
in the three-dimensional case.\label{fig:integration-domain}}
\end{figure}

As in the two dimensional case, it is possible to remove the correlations
between $dw^{b}$, $dw^{a}$, and $\phi$ via the coordinate change 

\begin{equation} \left\{ \begin{aligned} \alpha(x,y,z) = & x \\ \beta(x,y,z) = & \frac{\left(-\rho_{xy}x + y\right)}{\sqrt{1-\rho^2_{xy}}} \\ \gamma(x,y,z) = & \frac{\left[ \left(\rho_{xy}\rho_{yz}-\rho_{xz}\right)x + \left(\rho_{xy}\rho_{xz}-\rho_{yz}\right)y + (1-\rho_{xy}^2)  z \right]}{\sqrt{1-\rho^2_{xy}} \sqrt{1 - \rho_{xy}^2 - \rho_{xz}^2 - \rho_{yz}^2 + 2\rho_{xy}\rho_{xz}\rho_{yz}}}, \end{aligned} \right. \end{equation} yielding
the equation

\begin{equation}
P_{\alpha\alpha}+P_{\beta\beta}+P_{\gamma\gamma}=0.\label{eq:abc}
\end{equation}
Then we cast the problem into a convenient set of curvilinear coordinates
via the transformation

\begin{equation} \left\{ \begin{aligned} \alpha = & r \sin\theta \sin \varphi \\ \beta = & r \sin \theta \cos \varphi \\ \gamma = & r \cos \theta \end{aligned} \right. \longleftrightarrow \left\{ \begin{aligned} 	r = & \sqrt{\alpha^2 + \beta^2 + \gamma^2} \\ 	\theta = & \arccos\left( \frac{\gamma}{r} \right) \\ 	\varphi = & \textrm{arctan}\left( \frac{\alpha}{\beta} \right) \\ \end{aligned} \right. \end{equation} yielding
the modified problem

\begin{equation}
\frac{1}{\sin^{2}\theta}P_{\varphi\varphi}+\frac{1}{\sin\theta}\frac{\partial}{\partial\theta}\left(\sin\theta P_{\theta}\right)=0,\label{eq:phitheta}
\end{equation}
and the boundary condition for a near side trade becomes
\begin{equation}
P(0,\theta)=0,\quad P(\phimax,\theta)=0,\quad P(\varphi,\text{\textgreek{J}}(\varphi))=1.\label{eq:phitheta_trade_bc}
\end{equation}
The new integration domain after the introduction of spherical coordinates
is shown in Figure \ref{fig:integration-domain}. The problem (\ref{eq:phitheta})
can be further simplified by introducing one extra transformation,
\begin{equation}
\zeta=\ln\tan\theta/2,\label{eq:zeta_trans}
\end{equation}
 which changes the integration domain into the semi-infinite strip,
$0\le\phi\le\phimax,$ $\zeta\le Z\left(\phi\right)$, and the diffusion
equation (\ref{eq:phitheta}) into the form

\begin{equation}
P_{\varphi\varphi}+P_{\zeta\zeta}=0.\label{eq:phizeta}
\end{equation}
In this domain the near side trade boundary conditions become
\begin{equation}
P(0,\zeta)=0,\quad P(\phimax,\zeta)=0,\quad P(\varphi,Z(\varphi))=1.\label{eq:phizeta_trade_bc}
\end{equation}

The solution to problem (\ref{eq:phizeta}) that satisfies the first
two boundary conditions (\ref{eq:phizeta_trade_bc}) can be expressed
as a generalized Fourier series \cite{Lipton2013b}:

\begin{equation}
P(\varphi,\zeta)=\sum_{n=1}^{\infty}c_{n}\sin(k_{n}\varphi)e^{k_{n}\zeta}
\end{equation}
with $k_{n}=\frac{\pi n}{\phimax}$. The expansion coefficients $c_{n}$
can be determined by enforcing the third boundary condition in (\ref{eq:phizeta_trade_bc}).
To compute the coefficients, we introduce the integrals

\begin{equation}
J_{mn}=\intop_{0}^{\phimax}\sin(k_{m}\varphi)\sin(k_{n}\varphi)e^{(k_{n}+k_{m})Z(\varphi)}d\varphi
\end{equation}

\begin{equation}
I_{m}=\intop_{0}^{\phimax}\sin(k_{m}\varphi)e^{k_{m}Z(\varphi)}d\varphi
\end{equation}
Then the third boundary condition in (\ref{eq:phitheta_trade_bc})
becomes the matrix equation 
\begin{equation}
\sum_{n}J_{mn}c_{n}=I_{m},
\end{equation}
and the coefficients $c_{n}$ can be computed by matrix inversion
as $c=J^{-1}I$. 

\begin{figure}
\centering{}\includegraphics[scale=0.45]{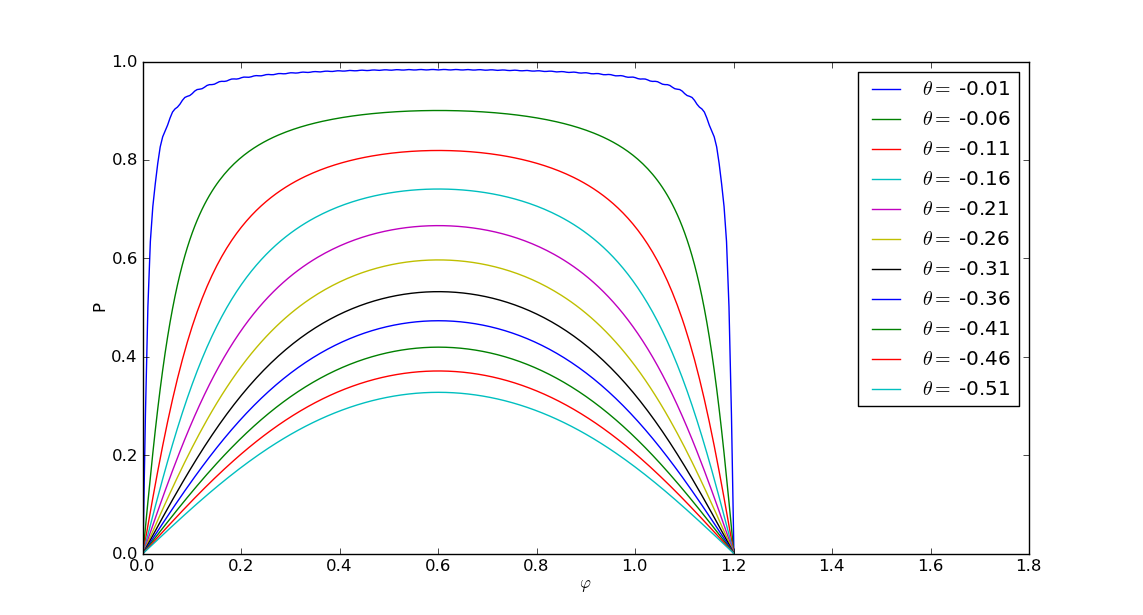}\caption{PDE solution to the problem (\ref{eq:phitheta}) in the $\varphi,\theta$
space, for $\phimax=1.2$ and boundary conditions as in (\ref{eq:phitheta_trade_bc})
.\label{fig:sol_profile}}
\end{figure}

When the boundary $\zeta=Z\left(\varphi\right)$ is approximately
linear, the integrals $I_{m}$ and $J_{mn}$ can be computed analytically.
Figure \ref{fig:sol_profile} shows the solution profile for the case
where the near side trade arrival process is uncorrelated from the
bid and ask queue sizes, i.e. $\rho_{xz}=\rho_{yz}=0$. In summary
we have a semi-analytic method of computing the probability $P\left(\varphi,\theta\right)$
or equivalently $P\left(x,y,z\right)$ for given initial values $x,y,z$
and corresponding correlations.

\section{Calibration\label{sec:Calibration}}

We proceed by calibrating the model jointly to the average mid-price
moves shown in Figure \ref{fig:cond_imbalance_emp-2} and to the empirical
probabilities of price moves and trade arrivals summarized in Figure
\ref{fig:cond_prob_empir-1}. It was shown in the previous section
that the model probabilities are determined by the book imbalance
(relative size of the bid and ask queues), the initial position of
the trade arrival process $\phi_{0}$ and the three elements of the
correlation matrix $\rho$. The book imbalance is a market observable,
and the remaining four parameters are to be estimated by the calibration
process. Model probabilities are computed semi-analytically as in
the previous section. Expectations of price changes up to the stopping
time defined by the arrival of a trade cannot be easily computed with
analytic methods and Monte Carlo simulations were used instead. 

\quad 

\begin{figure}[h]
\begin{centering}
\includegraphics[scale=0.4]{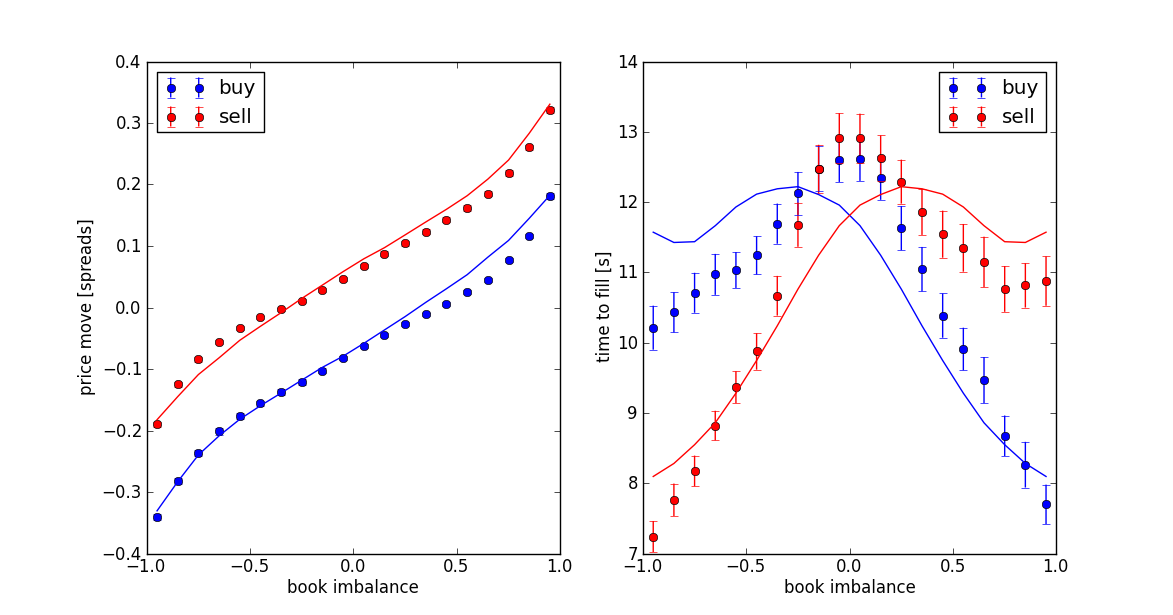}
\par\end{centering}

\caption{Empirical and model-derived average mid-price moves normalised by
the bid-ask spread and trade arrival times as a function of the book
imbalance. The calibrated correlation values are $\rho_{xz}=-\rho_{yz}=0.8$,
$\rho_{xy}=-0.1$ and $\phi_{0}=3.5$ $\mathrm{sec^{1/2}}$ \label{fig:cond_imbalance_emp-2-cal}}
\end{figure}

\begin{figure}[h]
\begin{centering}
\includegraphics[scale=0.6]{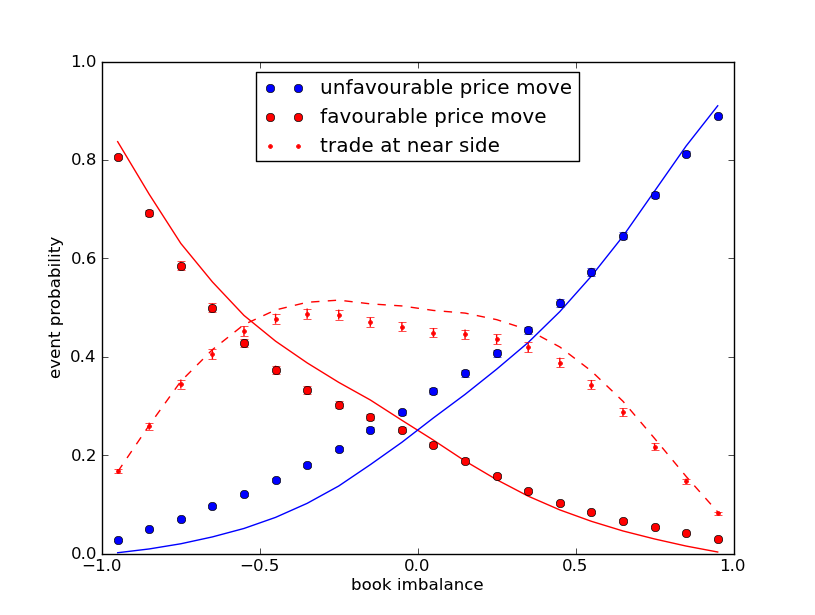}
\par\end{centering}

\caption{Empirical and calibrated model probabilities for the first occurrence
of a market event. The probability of obtaining a passive fill at
the current near side is shown by a dotted red line. The probability
that the price will move in favour of the broker while waiting for
the fill is shown by a solid red line and the probability that the
price will move against the broker by a solid blue line. \label{fig:Passive-fill-probabilities}}
\end{figure}

We calibrate the model against empirical probabilities of price movements
for equity VOD.L for all trading days in the first quarter of 2012.
The results are shown in Figures \ref{fig:cond_imbalance_emp-2-cal}
and \ref{fig:Passive-fill-probabilities}. Calibrations against data
from other liquid stocks displayed qualitatively similar results.
We note that the model reproduces the gap between the average mid-price
moves until the arrival of a buy or sell trade. The gap is controlled
by the correlation matrix $\mathbf{\rho}$ and in particular by the
two correlations between the trade arrival process and the bid and
ask queue processes. This shift in average price moves conditioned
on the side of the trade can be interpreted as the expected slippage
of a passive fill. When executing an order, a broker will typically
post a fraction of the total quantity at the near side in an attempt
to save part of the spread. However, when measuring the differences
in the average price achieved by aggressive and passive fills, we
observe that passive fills rarely achieve their expected savings of
one full spread. While part of this effect is generally attributed
to adverse selection (that is, the ability of other market participants
to take advantage of the timing given up by the broker when posting
at the near side), the slippage of orders posted at the near side
is also due to the interaction between queue depletion and trade flow,
as our model predicts. In the calibration described here, this effect
alone is responsible for the loss of about 60\% of the theoretical
spread captured by a passive fill. Figure \ref{fig:cond_imbalance_emp-2-cal}
also compares the empirical and model derived average arrival times
of a buy or sell trade as a function of book imbalance. The model
captures the gross features of the empirical shapes for moderate book
imbalance values, but is not as accurate in reproducing the steep
decrease of the arrival times at extreme book imbalance values.

Finally, we note that the model is able to reproduce the empirical
shapes of the event probabilities. As shown in Figure \ref{fig:Passive-fill-probabilities},
the probability of an unfavorable price movement increases as the
book gets heavier towards the near side, reaching almost 90\% in cases
of high imbalance. Knowing this, a broker can decide to keep the order
posted on the near side for moderate imbalance values and cross the
spread in a highly imbalanced book. Deriving such optimal spread crossing
policies is left for future work.

\section{Conclusions}

We analyzed the microstructure of trade arrival and its relationship
to the state of a limit order book. We observed empirically that the
arrival time of trades at the near side and the dynamics of the mid-price
until the arrival of a trade of a given side depend strongly on the
order book imbalance. We introduced a stochastic model with correlation
between the processes for the order queues at the top of the book
and a process representing the arrival of the trades at the near side
of the book. We were able to compute probabilities of price movement
and trade arrival in a semi-analytical form. This allowed us to perform
an efficient calibration of the model to empirical probabilities.
Since the model captures the dependence of the trade arrival probabilities
on the order imbalance, it can be used to construct optimal order
execution strategies.

\section*{Acknowledgments}

We would like to thank Ioana Savescu for many insightful discussions
and comments.

\bibliographystyle{plain}

\end{document}